\documentclass[5p,twocolumn,11pt]{elsarticle}

\usepackage{amsmath,amsfonts,amssymb,amsthm, booktabs, tabularx, placeins}

\usepackage[dvipsnames]{xcolor}
\usepackage{graphicx}
\usepackage{amsmath,amsfonts,amssymb}
\usepackage[breaklinks,colorlinks,urlcolor=Plum,citecolor=MidnightBlue,linkcolor=WildStrawberry]{hyperref}
\usepackage{enumitem}
\usepackage{orcidlink}
\usepackage{ulem}
\newcommand{\bea}{\begin{equation}\begin{aligned}} 
\newcommand{\eea}{\end{aligned}\end{equation}}
\newcommand{\be}{\begin{equation}}
\newcommand{\ee}{\end{equation}}

\newcommand{\msun}{M_{\odot}}
\newcommand{\td}{{\rm d}}

\biboptions{sort&compress}

\begin{document}

\title{{\bf The Dark Side of the Moon: \\
Listening to Scalar-Induced Gravitational Waves}}

\affiliation[0]{organization={Institut de Física d’Altes Energies (IFAE), The Barcelona Institute of Science and Technology},
    addressline={Campus UAB},
    city={Bellaterra (Barcelona)},
    postcode={08193},
    country={Spain}}

\affiliation[a]{organization={Institució Catalana de Recerca i Estudis Avançats (ICREA)},
    addressline={Passeig Lluís Companys 23},
    city={Barcelona},
    postcode={08010},
    country={Spain}}
\affiliation[J]{organization={Department of Physics, University of Wisconsin-Madison, Madison},
    addressline={WI 53706},
    city={Madison},
    country={USA}}
\affiliation[b]{organization={PRISMA$^+$ Cluster of Excellence \& Mainz Institute for Theoretical Physics (MITP), JGU Mainz, Germany}}
\affiliation[b2]{organization={Institut d’Astrophysique de Paris (IAP), CNRS, Sorbonne Université, FR-75014, France}}
\affiliation[d]{organization={Center for Astrophysics and Space Science (CASS), New York University Abu Dhabi},
    addressline={PO Box 129188},
    city={Abu Dhabi},
    country={UAE}}
\affiliation[c]{organization={Center for Astrophysics and Cosmology, University of Nova Gorica},
    addressline={Vipavska 13, POB 301},
    city={Nova Gorica},
    postcode={5000},
    country={Slovenia}}
\affiliation[e]{organization={Theoretical Physics Department, CERN},
    city={Geneva},
    postcode={1211},
    country={Switzerland}}
\affiliation[f]{organization={Physik-Institut, Universität Zürich},
    city={Zürich},
    postcode={8057},
    country={Switzerland}}

\affiliation[g]{organization={Departament de Física Quàntica i Astrofísica and Institut de Ciències del Cosmos (ICC), Universitat de Barcelona},
    addressline={Martí i Franquès 1},
    city={Barcelona},
    postcode={ES-08028},
    country={Spain}}

\author[0,a]{D. Blas\orcidlink{0000-0003-2646-0112}}
\author[J]{J. W. Foster\orcidlink{0000-0002-7399-2608}}
\author[b,b2]{Y. Gouttenoire\orcidlink{0000-0003-2225-6704}}
\author[d]{A. J. Iovino\orcidlink{0000-0002-8531-5962}}
\author[c]{\\I. Musco\orcidlink{0000-0003-1931-0642}}
\author[e,f]{ S. Trifinopoulos\orcidlink{0000-0002-0492-1144}}
\author[g]{M. Vanvlasselaer\orcidlink{0000-0002-8527-7011}}

\begin{abstract}
The collapse of large-amplitude primordial curvature perturbations into planetary-mass primordial black holes generates a scalar-induced gravitational wave background in the $\mu $Hz frequency range that may be detectable by future Lunar Laser Ranging and Satellite Laser Ranging data. We derive projected constraints on the primordial black hole population from a null detection of stochastic gravitational wave background by these experiments, including the impact of the standard model electroweak phase transition on the abundance of planetary-mass primordial black holes. We also discuss the connection between the obtained projected constraints and the recent microlensing observations by the HSC collaboration of the Andromeda Galaxy.
\\ \textit{Document Number: } CERN-TH-2026-021
\end{abstract}


\maketitle
\section{Introduction} 

In recent years, Primordial Black Holes (PBHs) have garnered renewed interest as they might explain some of the LIGO/Virgo/KAGRA observations and significantly contribute to the amount of dark matter today~(see for some recent reviews Resf.~\cite{Sasaki:2018dmp,Carr:2020gox,Green:2020jor,Escriva:2022duf,Carr:2023tpt,Carr:2026hot}). 
Among the various formation mechanisms proposed, the standard scenario involves the collapse of large-amplitude curvature perturbations generated during inflation, whose characteristic scale also determines the frequency of the associated Scalar-Induced Gravitational Wave background (SIGWs) through 
\begin{equation}
    f \sim 10^{-9}\,{\rm Hz}\,
    \left(\frac{M_{\rm PBH}}{M_\odot}\right)^{-1/2}.
\end{equation}
This connection between PBH mass and GW frequency enables the study of PBH formation across a wide mass range using different ongoing and future experiments\,\cite{Saito:2008jc}. 

Current Pulsar Timing Array (PTA) observations~\cite{NANOGrav:2023gor, NANOGrav:2023hde,EPTA:2023fyk, EPTA:2023sfo, EPTA:2023xxk,Reardon:2023gzh, Zic:2023gta, Reardon:2023zen,Xu:2023wog} are already sensitive to a potential SIGW signal in the nHz regime, expected from the formation of (sub)solar-mass PBHs, while the space-based interferometer LISA~\cite{LISACosmologyWorkingGroup:2022jok} will be able to explore the frequency band, i.e. mHz, associated with asteroid-mass PBHs. 
Therefore, these  mass ranges have been extensively investigated in the literature (see, e.g. Refs.~\cite{Bartolo:2018evs,Bartolo:2018rku,Cai:2018dig,DeLuca:2020agl,Kohri:2020qqd,Vaskonen:2020lbd,Franciolini:2023pbf,Liu:2023ymk,Balaji:2023ehk,Iovino:2024tyg,Franciolini:2023wjm,Domenech:2024rks,Gouttenoire:2025jxe,LISACosmologyWorkingGroup:2025vdz,Iovino:2025cdy}). 
Conversely, the planetary-mass window ($M_{\rm PBH}\sim[10^{-7},10^{-4}] \, M_\odot$) remains comparatively unexplored. 
Although such PBHs have long been considered in connection with microlensing observations~\cite{MACHO:2000qbb,Jedamzik:1998hc,Niikura:2019kqi,Niikura:2017zjd,Hawkins:2020zie,Gorton:2022fyb,Hawkins:2022vqo,Hawkins:2025mlo}, no dedicated experiment has yet been realized to specifically target SIGWs associated with this mass range. 
Several concepts were proposed a few years ago~\cite{Sesana:2019vho,Lu:2024yuo,Caliskan:2023cqm,Wang:2022sxn,Fedderke:2021kuy}, but none of them have progressed beyond the preliminary design or proposal stage.

More recently, Lunar Laser Ranging (LLR) and Satellite Laser Ranging (SLR)~\cite{Blas:2021mqw,Foster:2025nzf} have emerged as a promising avenue to explore this mass-frequency regime directly. The key idea is to exploit the resonant response of gravitationally bound systems, such as the Earth-Moon and Earth-satellite binaries, to long-wavelength GWs in the $\mu$Hz band. A stochastic GW background induces time-dependent perturbations of the orbital elements, which can be probed through high-precision laser-ranging measurements.

In this work, after briefly reviewing the formalism to compute the PBH abundance and the SIGWs in Section~\ref{sec:recap}, we briefly describe in Section\,\ref{sec:experiments}  the proposed experiments.
In Section\,\ref{sec:prospect} we constrain the amplitude and shape of the primordial curvature power spectrum, under the assumption that the SIGW background produced does not reach the future sensitivity of LLR, and some possible SLR missions.
We then discuss how these constraints can be translated into bounds on the PBH abundance, including also the effects of the Electroweak (EW) phase transition on the threshold for PBH formation, and the connection with existing PBHs microlensing limits and observations. 
Finally, we summarize our conclusions in Section~\ref{sec:conclusion}. 

\section{Review of PBH formation and SIGWs} 
\label{sec:recap}

In this section, we provide a brief summary of the main definitions and formulae relevant for the computation of the PBH abundance and the associated background of SIGWs. 
Throughout the paper, we assume, for simplicity, that the primordial curvature perturbation $\zeta$ follows a Gaussian distribution. Following a model-independent approach, we characterize the curvature power spectrum with a log-normal profile, defined by a peak scale $k_\star$, an amplitude $A$, and a width $\Delta$,  
\bea\label{eq:PSlog}
\mathcal{P}_{\zeta_G}(k)=\frac{A}{\sqrt{2\pi}\,\Delta}\,
\exp\!\left[-\frac{\ln^2(k/k_\star)}{2\Delta^2}\right].
\eea
We also assume that the Universe is radiation-dominated, except when accounting for modifications of the equation of state (EoS) in the calculation for the PBH abundance. 
Instead, in our SIGW computation, we neglect subleading effects such as the variation of the sound speed $c_{\rm s}$, across the EW and QCD transitions (see e.g. Refs.~\cite{Hajkarim:2019nbx,Abe:2020sqb,Franciolini:2023wjm})\footnote{The EW transition induces a few percent variation of the sound speed $c_s$, which modifies the radiation transfer functions entering the SIGW kernel only at a subleading level, leaving the spectrum essentially unaffected~\cite{Hajkarim:2019nbx, Abe:2020sqb}. This is in stark contrast with the PBH abundance, which is exponentially sensitive to $\mathcal{C}_{\rm th}$ and thus receives order-of-magnitude corrections from the same $\mathcal{O}(1\%)$ shift in the threshold.}.

\subsection{PBH formation: threshold statistics and peak theory}
Large-amplitude curvature perturbations could collapse into PBHs after re-entering the cosmological horizon (horizon crossing), i.e. when the wavelength of the perturbation is of the same size of the cosmological horizon. The resulting PBH mass follows the critical scaling relation~\cite{Choptuik:1992jv,Niemeyer:1997mt,Niemeyer:1999ak}
\be\label{eq:crit}
    M_{\rm PBH}(\mathcal{C}) 
    = \mathcal{K}\,M_k \left(\mathcal{C}-\mathcal{C}_{\rm th}\right)^{\gamma}\,,
\ee
where $\mathcal{C}$ is the peak of the compaction function measuring the perturbation amplitude, and the collapse occurs only if it exceeds the threshold $\mathcal{C}_{\rm th}$, whose value, as we will describe in the next section, depends on the power spectrum shape~\cite{Musco:2020jjb}.  The horizon mass at re-entry of the mode $k$ is given by
\be\label{eq:HorizonMasskH}
    M_k(t) \simeq 14\,M_\odot \left[\frac{k}{10^{6}\,{\rm Mpc}^{-1}} \right]^{-2}.
\ee
The amount of matter collapsing into PBHs of mass $M_{\rm PBH}$, measured at horizon crossing, is 
\be\label{eq:betak}
    \beta_k(M_{\rm PBH})
    = \int_{\mathcal{C}_{\rm th}}^\infty \!\td\mathcal{C}\; P_k(\mathcal{C})\,\frac{M_{\rm PBH}}{M_k}\,
    \delta\!\left[\ln\frac{M_{\rm PBH}}{M_{\rm PBH}(\mathcal{C})}\right],
\ee
where $P_k(\mathcal{C})$ is the probability distribution of $\mathcal{C}$. 
The present-day PBH mass function is then given by
\be\label{eq:df_PBH}
    f_{\rm PBH}(M_{\rm PBH})
= \frac{1}{\Omega_{\rm DM}}\int \frac{\td M_k}{M_k} \, \beta_k(M_{\rm PBH}) 
\left(\frac{M_{\rm eq}}{M_k}\right)^{1/2}\!,
\ee
with $M_{\rm eq}\simeq 2.8\times 10^{17}\,M_\odot$ the horizon mass at matter–radiation equality, and $\Omega_{\rm DM}=0.12h^{-2}$ the cold dark matter density~\cite{Planck:2018jri}.  

The precise determination of the PBH mass spectrum remains uncertain because different approximation schemes can be used, and there is no agreement  on which is the most consistent. 
The two most common formalisms are {\it threshold statistics} and {\it peak theory}, both of which can be recast in the form of Eq.~\eqref{eq:betak}, with different prescriptions for $P_k(\mathcal{C})$~\cite{Iovino:2024tyg,Gouttenoire:2025jxe}.  As we will demonstrate, our results are robust with respect to the method with which the PBH mass spectrum is computed.

\paragraph{Threshold statistics} 
Here, the probability is inferred from the statistics of the compaction function~\cite{Ferrante:2022mui,Gow:2022jfb}, defined as twice the mass excess relative to the areal radius. The PBH mass function in this framework reads~\cite{Young:2019yug,DeLuca:2019qsy}
\begin{equation}
\begin{aligned}
&f_{\rm PBH}(M_{\rm PBH}) = \frac{1}{\Omega_{\rm DM}}
\int_{M_k^{\rm min}}^{\infty}\frac{\td M_k}{M_k}
\left(\frac{M_{\rm eq}}{M_k}\right)^{1/2}
\left(\frac{M_{\rm PBH}}{\gamma M_k}\right) \\& \times
\left(\frac{M_{\rm PBH}}{\mathcal{K}M_k}\right)^{1/\gamma}\!
\frac{1}{\sqrt{2\pi}\,\sigma_c(M_k)\,\Lambda^{1/2}}
\exp\!\left[-\frac{8(1-\sqrt{\Lambda})^2}{9\sigma_c^2(M_k)}\right],
\end{aligned}
\end{equation}
where
\be
\Lambda = 1 - \left(\mathcal{C}_{\rm th} - 
\frac{3}{2}\left(\frac{M_{\rm PBH}}{\mathcal{K}M_k}\right)^{1/\gamma}\right).
\ee
The variance is given by
\be\label{eq:Sigma2}
\sigma_c^2(M_k) = \frac{16}{81}\int_{0}^{\infty}\frac{\td k}{k}\,(kr_m)^4W^2(kr_m)P_\zeta^{\mathcal{T}}(k,r_m),
\ee
where $P_\zeta^{\mathcal{T}}=T^2(k,r_m)P_\zeta(k)$, $W$ is the top-hat window function, and $T$ is the linear radiation  transfer function~\cite{Young:2022phe}.  

\paragraph{Peak theory} 
In this approach, PBHs correspond to sufficiently high peaks of the overdensity field~\cite{Yoo:2018kvb,Yoo:2019pma,Franciolini:2022tfm}. 
In the high-peak limit one finds
\begin{equation}
\begin{aligned}
f_{\mathrm{PBH}}(M_{\mathrm{PBH}})&= \frac{1}{\Omega_{\mathrm{DM}}} 
\int_{M_k^{\min}} \frac{\td M_k}{M_k}
\left(\frac{M_{\rm eq}}{M_k}\right)^{1/2}\left(\frac{2}{3}\right)^4 \\
&\times \frac{\mathcal{K}}{\gamma}
\frac{(1-\sqrt{\Lambda})^3}{\pi\,\sigma_c^4(M_k)\,(27\Lambda)^{1/2}}
\left(\frac{\sigma_{cc}(M_k)}{\sigma_c(M_k)}\right)^3 \\
&\times\left(\frac{M_{\rm PBH}}{\mathcal{K}M_k}\right)^{\frac{1+\gamma}{\gamma}} \exp\!\left[-\frac{8(1-\sqrt{\Lambda})^2}{9\sigma_c^2(M_k)}\right],
\end{aligned}
\end{equation}
where the rescaled moment is~\cite{Young:2014ana,Young:2020xmk}
\be
\sigma_{cc}^2(M_k)=\frac{16}{81}\int_0^\infty\frac{\td k}{k}(kr_m)^6W^2(kr_m)P_\zeta^{\mathcal{T}}(k,r_m).
\ee

\begin{figure*}[!t]
    \centering
    \includegraphics[width=0.99\textwidth]{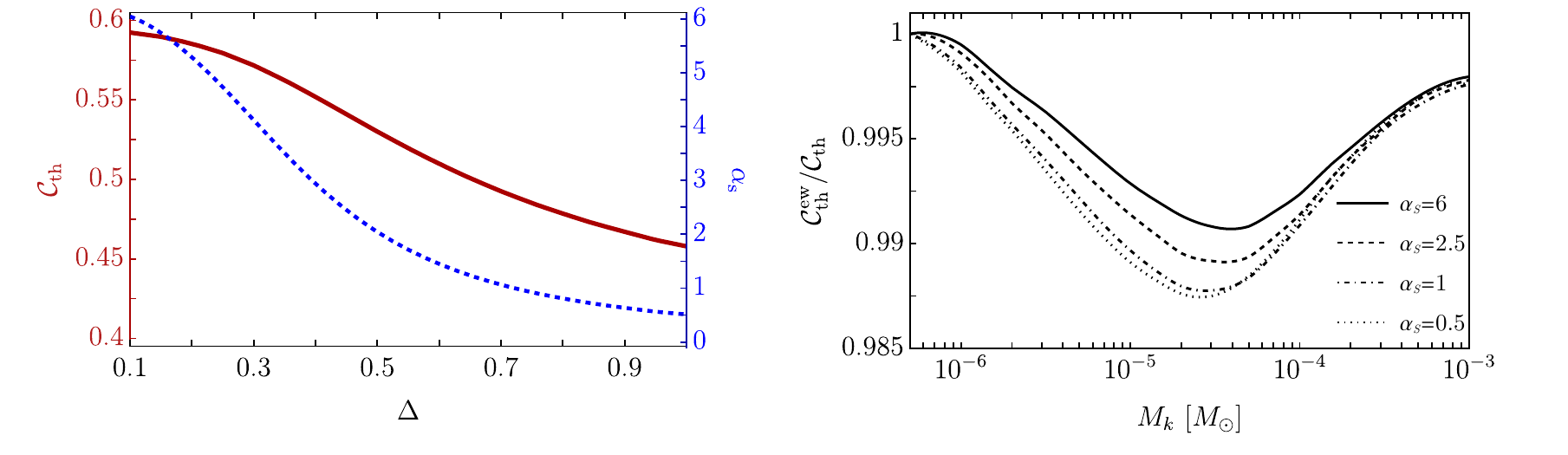}
    \caption{ \it \textbf{Left Panel}: Threshold $\mathcal{C}_{\rm th}$ (red solid) and shape parameter $\alpha_{\rm s}$ (blue dashed) values for different widths $\Delta$ of a log-normal power spectrum. \textbf{Right Panel}: The corresponding evolution of the threshold $\mathcal{C}^{\rm ew}_{\rm th}$ during the EW phase transition for different values of the shape parameter $\alpha_{\rm s}$, normalised with respect to the value of $\mathcal{C}_{\rm th}$ in the standard radiation dominated era.
 }
\label{fig:EWPT}
\end{figure*}

\subsection{PBH formation during the EW transition}

The gravitational collapse leading to the formation of a PBH occurs when the
maximum value of the compaction function, $\mathcal{C}(r_m)$, exceeds a critical threshold. This is measuring the perturbation amplitude for which there is a perfect equivalence between self-gravity and pressure gradients, corresponding to the asymptotic limit of a zero-mass black hole.  In general, the value of $\mathcal{C}_{\rm th}$ is not universal but depends on the shape profile of the cosmological perturbations collapsing into PBHs, i.e. the profile of $\zeta$, 
which is ultimately determined by the shape of the curvature power spectrum~\cite{Germani:2018jgr, Musco:2018rwt, Musco:2020jjb}.  
In this work we follow Refs.~\cite{Escriva:2019phb,Musco:2020jjb}, where a practical prescription is given to evaluate the collapse parameters directly from the curvature power spectrum.

For completeness, we summarize below the key steps required to compute the shape parameter $\alpha_{\rm s}$ and the corresponding value of threshold $\mathcal{C}_{\rm th}$.

\textit{i)} The position of the maximum of the compaction function, $r_m$, is obtained numerically by solving the condition
\begin{equation}
\int \frac{d k}{k} 
\left[
\left(k^2 r_m^2 - 1\right)\frac{\sin(k r_m)}{k r_m} + \cos(k r_m)
\right]
P_\zeta(k) = 0 .
\end{equation}
\textit{ii)} Once $r_m$ is determined, the shape parameter
\begin{equation}
\alpha_{\rm s}\equiv-C^{\prime\prime}(r_m)r^2_m/(4C(r_m))
\end{equation}
follows from the relation
\begin{equation}
[F(\alpha_{\rm s}) + F^2(\alpha_{\rm s})]\alpha_{\rm s}
= -
\left[
\frac{1}{2} + \frac{r_m}{2} \frac{\int d k\, k \cos(k r_m) P_\zeta(k)}
{\int d k\, \sin(k r_m) P_\zeta(k)}
\right],
\end{equation}
where, for convenience, we define
\begin{equation}
F(\alpha_{\rm s}) =
\left\{
1 -
\frac{(2/5)e^{-1/\alpha_{\rm s}}\alpha_{\rm s}^{1 - 5/(2\alpha_{\rm s})}}
{\Gamma\!\left(\frac{5}{2\alpha_{\rm s}}\right)
 - \Gamma\!\left(\frac{5}{2\alpha_{\rm s}}, \frac{1}{\alpha_{\rm s}}\right)}
\right\}^{1/2}.
\end{equation}
\textit{iii)} Finally, given $\alpha_{\rm s}$, the threshold for collapse in a pure radiation dominated universe can be approximated as~\cite{Escriva:2019phb}
\begin{equation}
\label{eq:C_th_final}
\mathcal{C}_{\rm th} \simeq 
\frac{4}{15} e^{-1/\alpha_{\rm s}}
\frac{\alpha_{\rm s}^{1 - 5/(2\alpha_{\rm s})}}
{\Gamma\!\left(\frac{5}{2\alpha_{\rm s}}\right)
 - \Gamma\!\left(\frac{5}{2\alpha_{\rm s}}, \frac{1}{\alpha_{\rm s}}\right)} .
\end{equation}
The left panel of Fig.~\ref{fig:EWPT} illustrates the dependence of the collapse threshold $\mathcal{C}_{\rm th}$ and of the shape parameter $\alpha_{\rm s}$ on the shape of the log-normal curvature power spectrum introduced in Eq.~\ref{eq:PSlog}.

The dynamics of PBH formation are very sensitive to the EoS of the primordial plasma,  showing features of critical collapse. In this context, the EoS plays a crucial role: PBH production is enhanced whenever the EoS softens, as it temporarily reduces the effect of the pressure against gravitational collapse.  

During the EW phase transition, changes in the effective number of relativistic degrees of freedom $g_\star$ and the associated variation in the sound speed $c_s$ modify the background expansion rate and the pressure gradients opposing the collapse. In particular, a transient softening of the EoS  around the EW scale reduces the resistance to gravitational collapse, lowering the threshold and enhancing the PBH formation probability during this epoch \cite{Jedamzik:1996mr,Jedamzik:1999am,Byrnes:2018clq,Carr:2019kxo,Sobrinho:2020cco,Franciolini:2022tfm,Escriva:2022bwe,Escriva:2020tak,Pritchard:2024vix,Pritchard:2025pcn,Musco:2023dak}.

In this paper, we extend the state-of-the-art of numerical simulations presented in Ref.~\cite{Musco:2023dak} to determine the threshold value $\mathcal{C}_{\rm th}$ during the EW phase transition, prolonging the previous analysis, focused on the QCD phase transition, to smaller horizon masses where the EW phase transition takes place. The results are shown in the right panel of Fig.~\ref{fig:EWPT}. The EW transition affects PBH masses in a range from $\sim 5\times10^{-7}\,\msun$ to $\sim 10^{-3}\,\msun$ when the Universe enters the QCD transition. We find that the correction on the threshold $\mathcal{C}_{\rm th}$ is modest, at the level of $\mathcal{O}(1\%)$, with the main correction  around $\sim [3,4]\times10^{-5}\,\msun$, where $c_s^2$ reaches the minimum value of $0.32$.\footnote{For comparison, QCD effects lead to significantly larger corrections on the threshold $\mathcal{C}_{\rm th}$, at the $\mathcal{O}(10\%)$ level, around $M_{\rm PBH}\sim3\,M_\odot$~\cite{Franciolini:2022tfm,Musco:2023dak} because of the larger softening of the EoS with respect to the EW transition.}. This small change on the threshold justifies the previous assumption of the mass distribution following critical collapse behavior of a radiation dominated universe, i.e. Eq.\,\ref{eq:crit}. In the numerical computations, while we fix $\gamma=0.36$ as in a radiation dominated universe, we take into account the shape dependence on $\mathcal{K}$, varying its value in the range  $\mathcal{K}\in[3.4,9]$ for $\alpha_{\rm s}\in[6,0.5]$, following the prescription of Ref.~\cite{Musco:2023dak}.

We observe, fully in agreement with Ref.~\cite{Musco:2023dak} where similar simulations were performed during the QCD phase transition, that broader profiles, corresponding to smaller values of $\alpha_s$, exhibit a stronger sensitivity to variations in the EoS and shift the main correction to smaller masses. More details on the EoS used in the numerical calculations to compute the threshold are given in the Appendix. 

\subsection{Scalar-induced gravitational waves}

Large scalar perturbations inevitably source tensor modes at second order around horizon re-entry~\cite{Matarrese:1997ay,  Matarrese:1992rp,Carbone:2004iv, Domenech:2020kqm, Bruni:1996im,Iovino:2025xkq}. This generates a stochastic background of SIGWs with the spectrum~\cite{Tomita:1975kj, Matarrese:1997ay,  Matarrese:1992rp}
\be\label{eq:GW}
h^2\Omega_{\rm GW}(k)=\frac{h^2\Omega_r}{24}
\frac{g_\star(T_k)}{g_\star^0}
\left(\frac{g_{\star s}(T_k)}{g_{\star s}^0}\right)^{-4/3}
\mathcal P_h(k),
\ee
where $g_{\star s}$ denotes the entropy degrees of freedom (evaluated at horizon crossing of $k$ and today, marked with $0$), while $h^2\Omega_r=4.2\times10^{-5}$ is the current radiation density fraction. 
The tensor power spectrum reads~\cite{Espinosa:2018eve,Kohri:2018awv}
\bea
\label{eq:P_h_ts}
\mathcal{P}_h(k)&=4\int_1^\infty \td t\int_0^1 \td s
\left[\frac{(t^2-1)(1-s^2)}{t^2-s^2}\right]^2\\
&\times{\cal I}_{t,s}^2 \mathcal{P}_\zeta\!\left(k\frac{t-s}{2}\right)
\mathcal{P}_\zeta\!\left(k\frac{t+s}{2}\right),
\eea
with the transfer function
\bea
\label{I_RD_osc_ave_ts}
{\cal I}_{t,s}^2 &= \frac{288(s^2+t^2-6)^2}{(t^2-s^2)^6}
\Bigg[\frac{\pi^2}{4}(s^2+t^2-6)^2\Theta(t-\sqrt{3}) \\
&\quad+\left(t^2-s^2-\tfrac{1}{2}(s^2+t^2-6)\ln\left|\frac{t^2-3}{3-s^2}\right|\right)^2\Bigg].
\eea

\section{LLR, eLO and eSLR setup}
\label{sec:experiments}

\begin{figure*}[!t]
    \centering
    \includegraphics[width=0.9\textwidth]{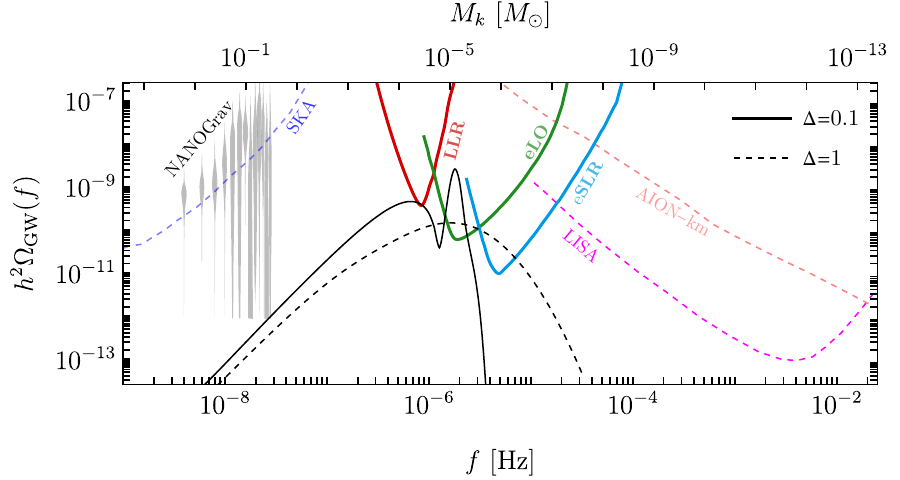}
    \caption{ \it Power-law-integrated sensitivity curves (solid lines) of LLR, eLO and eSLR~\cite{Foster:2025nzf}.
We also show the first 14 bins of NANOGrav 15 yrs experiment~\cite{NANOGrav:2023gor,NANOGrav:2023hvm} and future sensitivity (dashed lines) for planned experiments like SKA~\cite{Zhao:2013bba,Babak:2024yhu} and LISA~\cite{LISA:2022kgy} and AION-km~\cite{Badurina:2019hst,Badurina:2021rgt,Abdalla:2024sst}. For completeness, we also show in black the SIGW spectra for two benchmark scenarios, with $A=10^{-2}$, $k_{\star}=10^9$ Mpc$^{-1}$, $\Delta=0.1$ (solid) and $\Delta=1$ (dashed). }
\label{fig:situation}
\end{figure*}

In the LLR searches of GWs, the Earth-Moon system is treated as a resonant binary, whose orbital motion can be perturbed by long-wavelength GWs in the $\mu$Hz frequency band. A stochastic background of such waves induces a tidal acceleration on the relative separation vector with an oscillating pattern, thereby exciting harmonics of the orbital frequency and producing time-dependent perturbations in the osculating orbital elements that may accumulate with time, as well as in the round-trip light-travel time measured by LLR. Since the Earth-Moon baseline, i.e. the nominal orbital separation predicted by high-precision ephemerides, is known with exquisite accuracy, the perturbations from a very weak stochastic signal model are expected to be distinguishable from other effects~\cite{Blas:2021mqw,Foster:2025nzf,Foster:2025csl}.

In addition, we consider high-eccentricity orbits for laser-ranging that extend both current lunar and terrestrial capacities. 

$i)$ First, we study an eccentric lunar-orbit configuration (eLO), in which a satellite orbiting the Moon on a highly eccentric trajectory is tracked with a precision comparable to that of LLR. As done in Ref.~\cite{Foster:2025nzf}, we adopt orbital parameters corresponding to a period of 6.5 days and an eccentricity of 0.7. These parameters are representative of orbits currently planned for lunar navigation infrastructures such as \textit{Gateway}~\cite{nasaGatewayNASA}. While achieving the required laser-ranging precision for such missions would be challenging, this configuration provides a realistic benchmark for assessing the potential sensitivity of future lunar satellite tracking to long-wavelength GWs.

$ii)$ Second, we consider eccentric Satellite Laser Ranging (eSLR) scenarios involving artificial satellites on highly eccentric Earth orbits. In this case, the Earth-satellite system forms a gravitationally bound binary whose large eccentricity, like in the case of eLO, enhances the resonant response to long-wavelength GWs and results in amplified perturbations of the orbital elements. As an illustrative example, we use the orbit of the Magnetospheric Multiscale (MMS) mission, which was briefly laser ranged~\cite{Foster:2025nzf}. Although MMS was not designed as an SLR mission, and therefore does not provide the precision assumed in our projections, a future SLR experiment with similar orbital parameters is possible with current technologies. Other SLR missions which realize less eccentric and shorter-period orbits are also discussed in Ref.~\cite{Foster:2025nzf}.

A summary of the relevant GW sensitivities and the corresponding horizon mass ranges as a function of frequency is shown in Fig.~\ref{fig:situation}. Sensitivities via the binary resonance mechanism for the LLR, eLO, and eSLR scenarios are adopted from Ref.~\cite{Foster:2025csl} and depend in detail on the assumptions made in that work.

In the case of LLR, the projected sensitivities correspond to those achievable with 15 years of statistics-limited laser-ranging measurements and the assumption of the absence of degeneracies between the effects of a GW and other model parameters. For the satellite scenarios (eLO and eSLR), the corresponding missions do not exist, though the curves are based on ranging possibilities allowed by current technologies~\cite{Blas:2021mqw,Foster:2025nzf,Foster:2025csl}. Nonetheless, we anticipate that the prospects of detection of GWs from Ref.~\cite{Foster:2025nzf} will crystallize in a mission along this concept in the following years.

\section{Potential Constraints on PBH population}
\label{sec:prospect}

The non-detection of a stochastic SIGW background in the $\mu$Hz frequency band by future LLR, eLO and eSLR observations can be directly translated into constraints on the abundance of PBHs in the planetary-mass window. Since SIGWs are inevitably produced by the same large-amplitude curvature perturbations responsible for PBH formation, any experimental upper bound on $\Omega_{\rm GW}$ implies an upper limit on the amplitude of the primordial curvature power spectrum and, consequently, on the PBH mass function\footnote{We stress that assuming the observed signal to be entirely of SIGW origin is a conservative choice. If additional sources contributed to the signal, a smaller value of $A$ would be required, resulting in stronger bounds on PBH formation.}.

We develop projected sensitivities to $A$ as a function of $k_\star$ for SIGWs using the Fisher forecasting procedure for binary resonances developed in \cite{Foster:2025csl}. The procedure is as follows. In our log-normal spectrum model, the SIGW spectrum is fully determined  by $\{k_\star, A, \Delta\}$. At fixed values  of $k_\star$ and  $\Delta$, using the machinery of \cite{Foster:2025csl, Foster:2025nzf}, we compute the covariance for the binary resonance signal generated by that SIGW spectrum, which we denote $\mathbf{\Sigma}(A | k_\star, \Delta).$ Because the SIGW spectrum and therefore the signal covariance scale linearly with $A^2$, it is most convenient to define $\mathbf{\Sigma}( k_\star, \Delta) \equiv \mathbf{\Sigma}(A = 1|  k_\star, \Delta)$ so that 
\begin{equation}
\mathbf{\Sigma}(A |  k_\star, \Delta) = A^2 \mathbf{\Sigma}( k_\star, \Delta).
\end{equation}
The sensitivity to the squared amplitude  parameter $A^2$, denoted by $\sigma_{A^2}$ and the corresponding value of $A^2$ at which the signal-to-noise ratio is unity can then be calculated, assuming white noise measurement errors, by
\begin{equation}
    \sigma_{A^2} = \sqrt{\frac{2 \sigma^4}{\mathrm{Tr[\mathbf{\Sigma}(k_\star, \Delta)^2]}}}
\end{equation}
where $\sigma$ is the measurement precision. We adopt identical measurement precisions, observing cadences, and observing durations from \cite{Foster:2025nzf}. For concreteness, for LLR, the projected sensitivity corresponds to 15 years of statistics-limited measurements, with 260 light-travel-time measurements per year each at 3 mm precision (approximately 20 ps timing precision). For eLO (Gateway orbit: period 6.5 days, eccentricity 0.7) and eSLR (MMS-like orbit: eccentricity 0.9084), we assume 10-year campaigns with a ranging precision ten times larger than that of LLR. Full details on these choices, including the discussion of the measurement cadence and the adopted orbital parameters for each scenario, are provided in Appendix C of Ref.\,\cite{Foster:2025nzf}. The expected sensitivity to $A^2$ can then be trivially related to the expected sensitivity to $A$.

\begin{figure}[!t]
    \centering
    \includegraphics[width=0.49\textwidth]{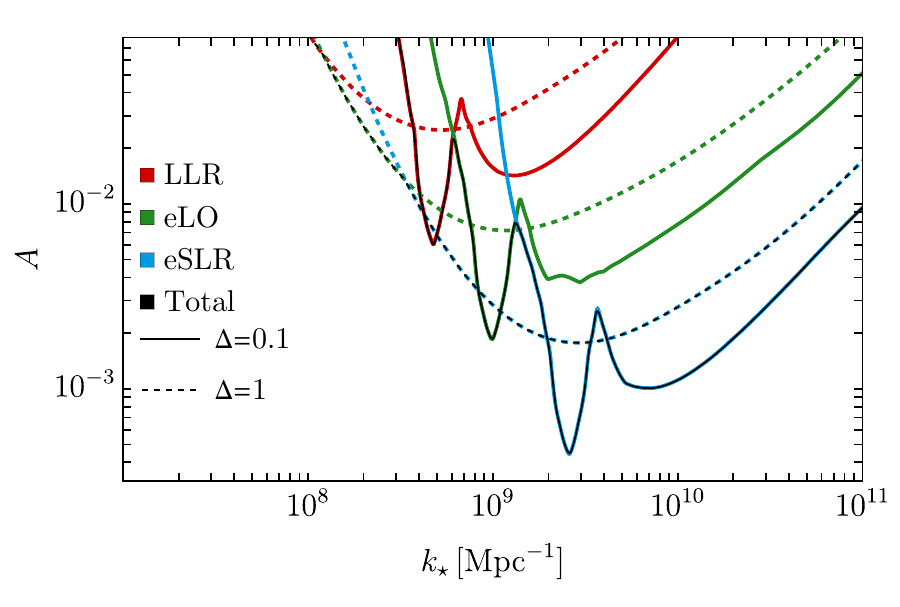}
    \caption{ \it  Prospects from LLR, eLO and eSLR experiments based on the predicted sensitivity, which provide an upper bound on the power spectrum amplitude $A$ in case no signal is observed for different spectral shape $\Delta$ in the case of a log-normal power spectrum.} 
\label{fig:moneyplot1}
\end{figure}

We evaluate our projected sensitivity to $A$ in the $\Delta =  0.1$ and $\Delta = 1.0$ scenarios at 1000 log-spaced values between values of $k_\star$ between $10^{7}\,\mathrm{Mpc}^{-1}$ and $10^{12}\,\mathrm{Mpc}^{-1}$. The results are shown in Fig.\,\ref{fig:moneyplot1}. For narrow power spectra, $\Delta=0.1$, the SIGW spectrum exhibits a characteristic double-peak structure~\cite{Pi:2020otn}, which translates into two local minima in the allowed amplitude $A$ for each individual experiment (see also for instance Fig.\,\ref{fig:situation}). 
When combining all experiments, the resulting constraint (either solid or dashed black lines) displays four local minima: three associated with the dominant sensitivity ranges of LLR, eLO and eSLR, and an additional feature at large $k_\star$ induced by the secondary peak in the eSLR case. 
This extra minimum disappears for broader spectra, i.e $\Delta\gtrsim0.5$\,\cite{Pi:2020otn}. As a consequence, the second case with $\Delta=1$ has a SIGW spectrum smoother and then the constraints are correspondingly relaxed. When combining the sensitivities, because one observation typically dominates the expected sensitivity, rather than projecting a joint analysis, we merely take the maximal sensitivity envelope for simplicity.

\begin{figure*}[!t]
    \centering
    \includegraphics[width=0.96\textwidth]{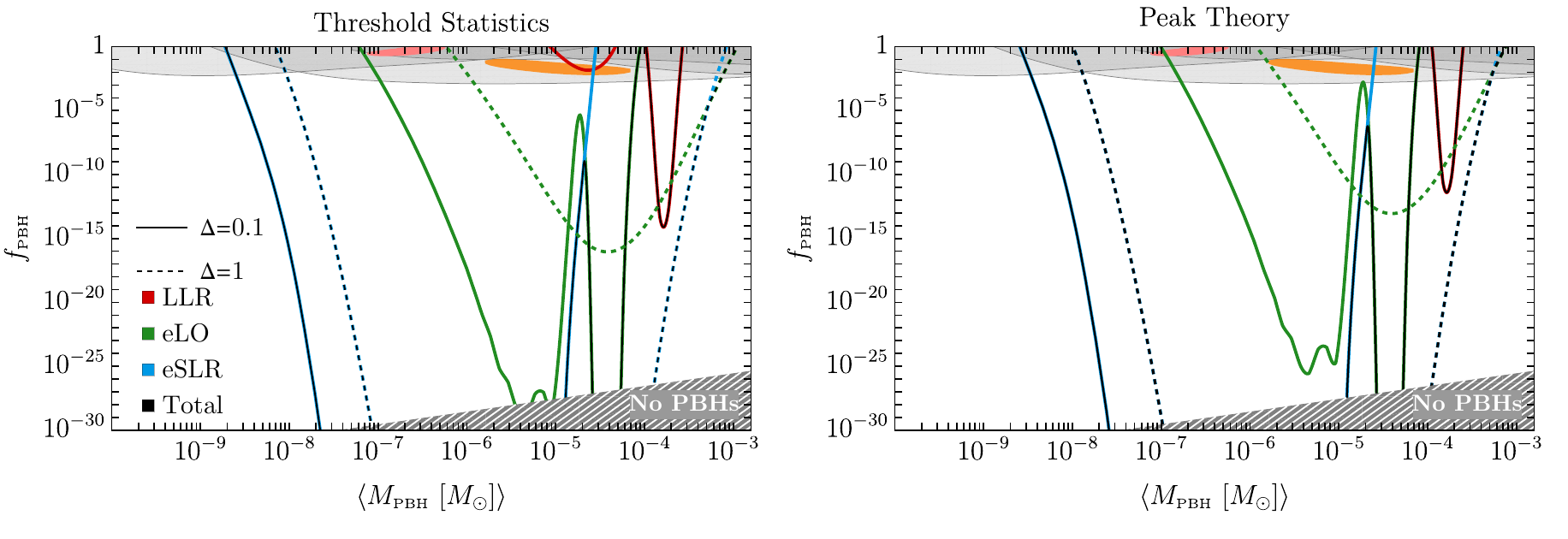}
    \caption{ \it \textbf{Left Panel}: The implied upper limits on PBH abundance using the threshold statistics formalism. The orange and pink regions are respectively the 95\% CL allowed region of PBH abundance, obtained by 
microlensing events in the OGLE+HSC\,\cite{Niikura:2017zjd,Mroz:2017mvf,Niikura:2019kqi} (orange) (see also~\cite{Sugiyama:2021xqg}) and in the HSC \,\cite{Sugiyama:2026kpv} (pink)  data due to the potential existence of PBHs. \textbf{Right Panel}: As in the left panel,  but with PBH abundances calculated via the peak theory formalism. } 
\label{fig:moneyplot2}
\end{figure*}

The constraints in the $(k_\star,A)$ plane can be mapped onto the $(\langle M_{\rm PBH}\rangle,f_{\rm PBH})$ plane by computing the PBH abundance assuming Gaussian primordial fluctuations. The resulting bounds are shown in Fig.~\ref{fig:moneyplot2}, where the PBH abundance is computed using threshold statistics (left panel) and peak theory (right panel), respectively. Existing observational constraints are also shown in gray, including EROS~\cite{EROS-2:2006ryy}, OGLEhc~\cite{Mroz:2024wia}, OGLE~\cite{Mroz:2024mse,Mroz:2024wag} and HSC~\cite{Niikura:2017zjd}. The shaded regions at the bottom of the panels indicate parameter choices for which fewer than one PBH is expected within the present Hubble volume. This condition is given by~\cite{Carr:1997cn,Carr:2020erq}
\be
    f_{\rm PBH}\, 4\pi \Omega_{\rm DM} \frac{M_{\rm Pl}^2}{H_0}
    > \langle M_{\rm PBH} \rangle \, ,
\ee
below which PBHs become observationally irrelevant.
The colored contours correspond to regions of the parameter space for which the microlensing events registered by OGLE+HSC\,\cite{Niikura:2017zjd,Mroz:2017mvf,Niikura:2019kqi} (orange color) (see also~\cite{Sugiyama:2021xqg}) and HSC\,\cite{Sugiyama:2026kpv} (pink color) can be of primordial origin\footnote{It is possible that the candidate microlensing events identified in Ref.\,\cite{Sugiyama:2026kpv} are false positives or are all the events are due to lenses in the Milky Way disk rather than PBHs. A possible primordial origin of these microlensing events can be explained in the axionlike curvaton models\,\cite{Kasai:2026yna}.}.

Assuming Gaussian primordial fluctuations, we find that a null detection of SIGWs by all three proposed experiments would lead to very stringent constraints on the PBH population. For narrow spectra, $\Delta=0.1$, PBHs are excluded over the entire mass range $\sim[2\times10^{-8},5\times10^{-5}]\,M_\odot$, independently of the PBH abundance formalism used. A narrow window around $\langle M_{\rm PBH}\rangle \simeq 2\times10^{-5}\,M_\odot$ remains formally allowed, but only at abundances $f_{\rm PBH}\lesssim10^{-10}$, rendering PBHs negligible as a dark matter component. For broader spectra, $\Delta=1$, the exclusion region relaxes at lighter masses, but PBHs are still ruled out in the interval $\sim[10^{-7},10^{-4}]\,M_\odot$. Remarkably, a non-detection by eLO could fully exclude the possibility that the events detected by OGLE+HSC\,\cite{Niikura:2017zjd,Mroz:2017mvf,Niikura:2019kqi,Sugiyama:2021xqg} (orange region) are of primordial origin; while in the case of broad spectrum only eSLR can fully exclude the same possibility for the recent claim presented by HSC\,\cite{Sugiyama:2026kpv} (pink region).
We stress that all exclusion regions quoted below refer specifically to the log-normal profile of Eq.~\eqref{eq:PSlog}; constraints derived under different spectral templates would in general differ quantitatively, though the qualitative picture is expected to remain robust.

Taking into account individual experiments, the constraints at the lowest PBH masses are systematically dominated by eSLR, largely independent of the spectral width. In particular, around $\langle M_{\rm PBH}\rangle \sim 10^{-5}\,M_\odot$, corresponding to SIGW frequencies most sensitive to the EW phase transition, the bounds are entirely driven by eSLR. 
In this mass range, the temporary softening of the equation of state during the EW phase transition mildly enhances PBH formation, leading to a slightly more stringent upper limit on $f_{\rm PBH}$. At larger masses, $\langle M_{\rm PBH}\rangle \sim 10^{-4}\,M_\odot$, the constraints are instead dominated by eLO, while those from LLR are found to be subdominant or negligible compared to existing microlensing bounds. For narrow spectra, the dominance alternates between LLR and eLO, reflecting the interplay between the sharp SIGW spectral features and the experimental sensitivity curves.

We can provide a simple argument to estimate the observation time needed to probe the region of parameter space associated with the recent microlensing observations by HSC\,\cite{Sugiyama:2026kpv}. As shown in Figure~\ref{fig:moneyplot2}, in order to probe this region one requires $f_\text{PBH} \sim [10^{-3}, 1]$, which in turn corresponds to a primordial power spectrum amplitude of order $A \sim 10^{-2}$ when using the threshold statistics formalism, and to an amplitude approximately 1.5 times smaller, $A \sim 6\times 10^{-3}$, when using the peak theory formalism. Both values are roughly one order of magnitude larger than the benchmark sensitivity 
$A \sim 10^{-3}$ reached by eSLR in the relevant range, i.e. $k\sim10^{10}$ Mpc$^{-1}$, the most constraining experiment in this region after 10 years of observation. To estimate how much observation time would be needed to reach such amplitudes, we exploit the scaling of the sensitivity with observation time derived in Refs.~\cite{Foster:2025csl,Foster:2025nzf}. 
For a broadband stochastic gravitational wave background, the Fisher information  scales as $\mathcal{I} \propto t_\text{obs}^8 / (\sigma^4 \Delta t^2)$, implying 
that the sensitivity to the energy density $\Omega_\text{GW} \propto A^2$ scales 
as $t_\text{obs}^4/\sigma^2$. Consequently, the sensitivity to the amplitude 
$A$ scales as$
    \sigma_A \propto \sigma/t_\text{obs}^2.$
Using this scaling, the observation time $t_2$ required to achieve a sensitivity 
$A_2$ can be related to the fiducial time $t_1 = 10$\,yr (at which 
$A_1 \sim 10^{-3}$ is reached) by
\begin{equation}
    \frac{A_1}{A_2} = \left(\frac{t_2}{t_1}\right)^2
    \quad \Longrightarrow \quad
    t_2 = t_1 \sqrt{\frac{A_1}{A_2}}.
\end{equation}
Applying this relation to the two formalisms:
\begin{itemize}
    \item \textit{Threshold statistics} ($A_2 \sim 10^{-2}$):
    \begin{equation}
        t_2 = 10\,\text{yr} \times \sqrt{\frac{10^{-3}}{10^{-2}}} 
             \approx 3\,\text{yr};
    \end{equation}
    \item \textit{Peak theory} ($A_2 \sim 6\times 10^{-3}$):
    \begin{equation}
        t_2 = 10\,\text{yr} \times \sqrt{\frac{10^{-3}}{6\times 10^{-3}}} 
             \approx 4\,\text{yr}.
    \end{equation}
\end{itemize}
These rough estimates suggest that approximately 3-4 years of eSLR observations would already be sufficient to probe the region associated with the microlensing events, using technology that is already available today.
\section{Conclusions}
\label{sec:conclusion}
In this work, we computed the projected constraints on the amplitude of the curvature power spectrum, requiring that the amplitude of the resulting SIGW background does not exceed the future sensitivity of the proposed LLR, eLO and eSLR experiments. When PBHs form through the collapse of large curvature perturbations, these bounds on SIGWs can be directly translated into upper limits on the PBH abundance. We derived these limits in different scenarios, estimating the PBH population using both threshold statistics and peak theory. Conversely, a detection of a SIGW signal in the $\mu$Hz band would provide compelling evidence for PBH formation processes tied to early-Universe physics, with particular sensitivity to epochs such as the EW phase transition.

Our results show that, in the absence of a detected SIGW signal, future laser-ranging experiments have the potential to severely constrain the existence of PBHs in the planetary-mass window. For narrow curvature power spectra, PBHs are excluded over essentially the entire mass range $M_{\rm PBH}\sim[10^{-8},10^{-5}]\,M_\odot$, independently of the PBH formation formalism, with only a narrow mass interval remaining allowed at abundances so small that PBHs are observationally irrelevant. Instead, for broader spectra, the constraints exclude the entire mass range $M_{\rm PBH}\sim[10^{-7},10^{-4}]\,M_\odot$.

Around $M\sim10^{-5}\,\msun$, corresponding to SIGW frequencies affected by the electroweak phase transition, the projected bounds are dominated by eSLR and remain strong enough to push the allowed PBH abundance below the level corresponding to fewer than one PBH within the present Hubble volume. At larger masses, the constraints are instead driven by eLO, while those from LLR are found to be subdominant or comparable to existing microlensing bounds, depending on the spectral width. Moreover, a non-detection from these experiments would exclude the possibility that candidate events seen by several microlensing experiments\,\cite{Niikura:2017zjd,Mroz:2017mvf,Niikura:2019kqi,Sugiyama:2021xqg,Sugiyama:2026kpv} are of primordial origin.

Although these constraints are indirect, as they arise from the non-detection of SIGWs rather than from direct PBH observations, if realized, they would represent the leading bounds on the abundance of planetary-mass PBHs over a large region of parameter space. As such, they provide a powerful complement to microlensing searches and establish laser-ranging experiments as a novel probe of early-Universe physics, particularly of PBH formation during epochs in which the EoS of the primordial plasma deviates from the radiation dominated Universe, as during the EW phase transition.

Several extensions of this work are worth exploring. A first direction concerns extending the analysis to very broad power spectra for which the computation of the PBH abundance is still under debate\,\cite{Ianniccari:2024bkh,Fumagalli:2024kgg} and additional effect, that alter the original mass function, such as clusterogenesis, can happens\,\cite{deluca:2022bjs,Iovino:2026vhv}. Further it is important to include primordial non-Gaussianities, which can substantially affect both the PBH abundance and the associated scalar-induced gravitational-wave background~\cite{Iovino:2024tyg,Iovino:2025cdy,Gouttenoire:2025jxe}. Moreover, our constraints do not apply for PBH formed from mechanisms beyond the standard collapse of large curvature perturbations. These include collapses induced by attractive long range force~\cite{Kouvaris:2013kra,Kouvaris:2018wnh,Acevedo:2020gro,Chang:2018bgx,Gross:2021qgx,Kawana:2021tde,Lu:2024xnb,Ralegankar:2024zjd,Flores:2020drq,Flores:2021jas,Domenech:2023afs,Flores:2023zpf}, late-blooming during first-order phase transitions~\cite{Kodama:1982sf,Hsu:1990fg,Baker:2021nyl,Baker:2021sno,Liu:2021svg,Hashino:2021qoq,Kawana:2022olo,Lewicki:2023ioy,Gouttenoire:2023naa,Baldes:2023rqv,Gouttenoire:2023bqy,Salvio:2023ynn,Gouttenoire:2023pxh,Jinno:2023vnr,Flores:2024lng,Lewicki:2024ghw,Lewicki:2024sfw,Cai:2024nln,Ai:2024cka,Murai:2025hse,Arteaga:2024vde,Banerjee:2024cwv,Hashino:2025fse,Franciolini:2025ztf}, late-annihilation of domain-wall networks~\cite{Ferrer:2018uiu,Gelmini:2023ngs,Gelmini:2022nim,Gouttenoire:2023gbn,Gouttenoire:2023ftk,Ferreira:2024eru,Lu:2024ngi,Gouttenoire:2025ofv}, domain-wall networks bounded by cosmic strings~\cite{Ge:2019ihf,Ge:2023rrq,Dunsky:2024zdo}, spherical domains generated during inflation~\cite{Garriga:2015fdk,Deng:2016vzb,Deng:2017uwc,Deng:2020mds,Kusenko:2020pcg,Maeso:2021xvl,Escriva:2023uko,He:2023yvl,Huang:2023mwy,Kitajima:2020kig,Kasai:2023ofh,Kasai:2023qic} or from eternally inflating regions\,\cite{Franciolini:2026ujb}.

Finally, we note that the sensitivity projections presented in this work are optimistic by construction, as they assume statistics-limited measurements in the absence of a dedicated noise analysis. Once actual data from future laser-ranging missions become available, a dedicated analysis including a realistic noise budget and systematic effects will be required to fully assess the constraining power of these experiments on the PBH population.

\section*{Acknowledgments}
We thank G. Perna and A. Riotto for useful discussions. We thank also the Institute for Fundamental Physics of the Universe (IFPU) in Trieste  for hosting the Focus Week ``PBHs in the Multimessenger Era" in November 2025 where this collaboration has been initiated. 
Y.G. acknowledges support by the Cluster of Excellence ``PRISMA+'' funded by the German Research Foundation (DFG) within the German Excellence Strategy (Project No. 390831469). M.V. is funded by the European Union (ERC, HoloGW, Grant Agreement No. 101141909). M.V. also acknowledges financial support from Grant CEX2024-001451-M funded by MICIU/AEI/10.13039/501100011033, from Grant No. PID2022-136224NB-C22 from the Spanish Ministry of Science, Innovation and Universities, and from Grant No. 2021-SGR-872 funded by the Catalan Government. S.T. is supported by the Swiss National Science Foundation - project n. P5R5PT\_222350, and acknowledges the CERN TH Department for hospitality while this research was being carried out. This project has also received funding from the European Union’s Horizon Europe research and innovation programme under the Marie Skłodowska-Curie Staff Exchange grant agreement No 101086085 - ASYMMETRY. I.M. has received support from the European Union’s Horizon Europe research and innovation program under the Marie Sklodowska-Curie COFUND postdoctoral programme grant agreement No. 101081355-SMASH and from the Republic of Slovenia and the European Union from the European Regional Development Fund. Views and opinions expressed are, however, those of the authors only and do not necessarily reflect those of the European Union or the European Research Council. Neither the European Union nor the granting authority can be held responsible for them.

This publication is part of the grant PID2023-146686NB-C31 funded by MICIU/AEI/10.13039/501100011033/ and by FEDER, UE.
IFAE is partially funded by the CERCA program of the Generalitat de Catalunya.
This publication is part of the grant CEX2024-001441-S funded by MICIU/AEI/10.13039/501100011033/.
This work is supported by ERC grant (GravNet, ERC-2024-SyG 101167211, DOI: 10.3030/101167211).  
D.B. acknowledges the support from the European Research Area (ERA) via the UNDARK project of the Widening participation and spreading excellence programme (project number 101159929).  Project supported by a 2024 Leonardo Grant for Scientific Research and Cultural Creation from the BBVA Foundation. The BBVA Foundation accepts no responsibility for the opinions, statements and contents included in the project and/or the results thereof, which are entirely the responsibility of the authors. 
\section*{\small Appendix: Equation of State in the Early Universe}
\label{app:EoS}
The matter content of the early Universe after the end of inflation is made by relativistic matter described by the EoS $p=\rho/3$, where the pressure $p$ is proportional to $\rho$, the total energy density of the fluid. This simple form of the EoS, is obtained in the approximation of a perfect fluid (no shear, no viscosity), when the rest mass density is negligible with respect to the kinetic component. Afterwards the Universe is matter dominated ($p=0$), i.e. the same perfect fluid approximation in the opposite limit, although strictly speaking this holds only for the background solution and at linear order of cosmological perturbations, while during the formation of structures, like stars and galaxies, a more sophisticated EoS needs to be taken into account, including the velocity dispersion of particles.

During the radiation dominated epoch, the temperature decreases with cosmic expansion with matter going through several transitions, characterized by a non-negligible softening of the equation of state, introducing an intrinsic scale into the problem. These include the electroweak transition at temperature $T\sim 100\,$GeV, periods of quark annihilation, the QCD confinement transition at $T\sim 100\,$MeV and the primordial nucleosynthesis with $e^+e^-$ annihilation at $T\sim 500\,$keV.

Here we are interested in the formalism that can be applied to the EW and QCD transitions, assuming for simplicity the chemical potential $\mu=0$ as in the calculations of~\cite{Borsanyi:2016ksw,Bhattacharya:2014ara}, describing a transition in equilibrium. This is an excellent approximation in the context of the early Universe due to the smallness of the cosmic baryon-to-photon ratio, which gives $\mu \sim 10^{-10}$.
\begin{figure*}[!t]
    \centering
    \includegraphics[width=0.99\textwidth]{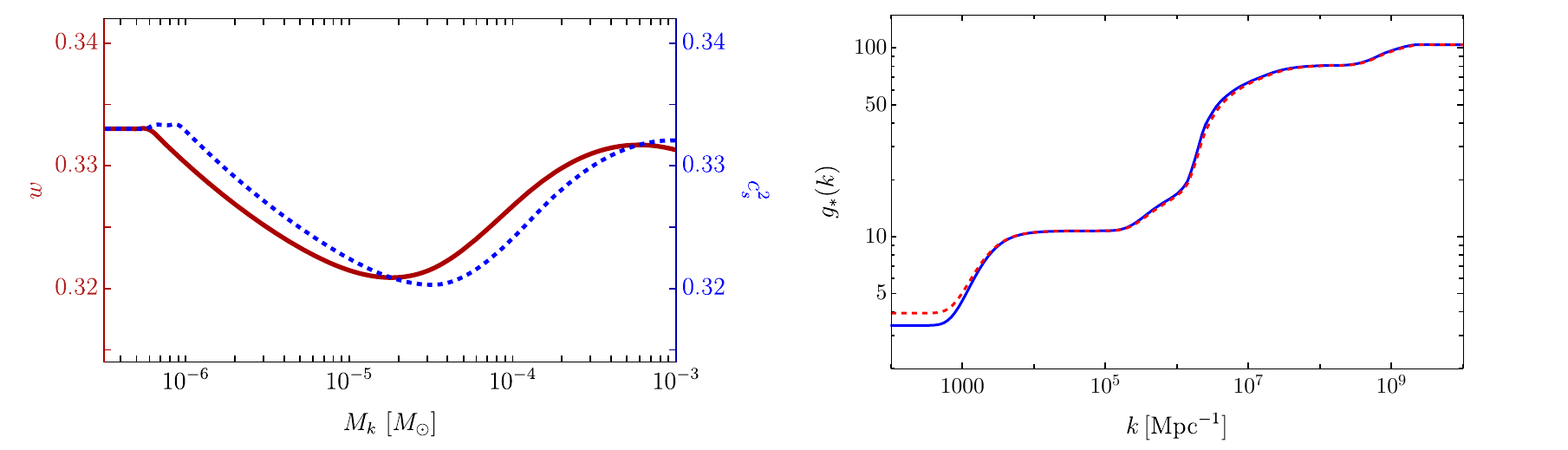}
    \caption{ \it \label{fig:EoS} \textbf{Left panel}: Evolution of the squared sound speed, $c_s^2$, and the equation of state parameter $w(T)\equiv p/\rho$ during the EW phase transition as a function of the cosmological horizon mass $M_{ k}$ in units of a solar mass. \textbf{Right panel}: \textcolor{blue}{Evolution of the degrees of freedom, $g_*$, and entropy degrees of freedom, $g_{s*}$, as a function of $k$.}}
\end{figure*}

\subsection*{\bf \small{The QCD and the electroweak transition}}
Using detailed lattice gauge calculations, considering
realistic quark masses, one can compute the EoS at zero chemical potential for the QCD transition in the early Universe. 
The ratio $w$ between pressure $p$ and total energy density $\rho$ of the medium is given by
\begin{equation}
w(T) \equiv \frac{p}{\rho} = \frac{4g_{\star s}(T)}{3g_{\star}(T)} - 1\, ,
\end{equation}
where the functions $g_{\star}(T)$ and $g_{\star s}(T)$ are defined by 
\begin{equation}
   g_{\star}(T) \equiv \frac{30\rho}{\pi^2T^4} 
    \quad \textrm{and} \quad 
   g_{\star s}(T) \equiv \frac{45s}{2\pi^2T^3}\, ,
\end{equation}
with $s$ being the entropy density. The square of the speed of sound
$c_s^2 \equiv \partial p/\partial \rho |_s$
may be computed as
\begin{equation}
    c_s^2(T) = \frac{4(4g_{\star s} + Tg_{\star s}^{\prime})}
                    {3(4g_{\star} + Tg_{\star}^{\prime})} - 1\, ,
    \label{eq:EoS}
\end{equation} 
where a prime denotes a derivative with respect to temperature.

These lattice calculations show clearly that the QCD quark-to-hadron 
transition is not a phase transition but a cross-over. 
Ref.~\cite{Borsanyi:2016ksw} in particular has added to these calculations results from the literature concerning the electroweak transition to provide the cosmic equation of state between $T \simeq 280\,$GeV and $T \simeq 1\,$MeV. 

Fig.~\ref{fig:EoS} shows the evolution of the sound speed squared, $c_s^2$, and of the equation-of-state parameter $w(T)\equiv p/\rho$ across the electroweak (EW) phase transition. The horizontal axis is expressed in terms of the cosmological horizon mass $M_k$, normalized to the solar mass $M_\odot$. In this parametrization, effects from the EW transition occur in the range $M_k \simeq 5\times10^{-7}\,M_\odot$ to $M_k \simeq 10^{-3}\,M_\odot$, after which effects from the QCD transition become dominant, see e.g.~\cite{Franciolini:2022tfm}. The full evolution of the equation of state, including the QCD transition and the epoch of $e^+e^-$ annihilation, is shown in Fig.~1 of Ref.~\cite{Musco:2023dak}.
\bibliographystyle{elsarticle-num} 
\bibliography{ref}

\newpage

\appendix

\end{document}